\documentclass{phb-proc4-auth}

\usepackage{graphicx}
\usepackage{amssymb}

\begin{document}
\begin{frontmatter}

\journal{SCES '04}

\title{Intermediate valence behavior in CeCo$_9$Si$_4$}

\author[1]{M. El-Hagary}
\author[1]{H. Michor}
\author[1]{E. Bauer}
\author[1]{R. Gr\"ossinger}
\author[2]{P. Kerschl}
\author[2]{D. Eckert}
\author[2]{K.-H. M\"uller}
\author[3]{P. Rogl}
\author[4]{G. Giester}
\author[1]{G. Hilscher\corauthref{*}}

\address[1]{Institut f\"ur Festk\"orperphysik, T.U. Wien, A-1040 Wien, Austria}
\address[2]{Leibniz-Institut f\"ur Festk\"oper- und 
Werkstoffforschung, IFW Dresden, D-01171 Dresden, Germany}
\address[3]{Institut f\"ur Physikalische Chemie, Universit\"at Wien, A-1090 Wien, Austria}
\address[4]{Institut f\"ur Mineralogie und Kristallographie, Universit\"at Wien,
A-1090 Wien, Austria}

\thanks[A]{The work was supported by the Austrian Science Foundation Fonds
under project P-15066-Phy; PK is supported by the BMBF project No. 03SC5DRE.
M. El-H. is on leave of Helwan University, Cairo, Egypt.}

\corauth[*]{Corresponding Author: 
Phone: (+431) 58801-13130 
Fax: (+431) 58801-13199, Email: hilscher@ifp.tuwien.ac.at}

\begin{abstract}

The novel ternary compound CeCo$_9$Si$_4$ has been studied by means of specific heat, magnetisation, and transport 
measurements. 
Single crystal X-ray Rietveld refinements reveal a fully ordered distribution of Ce, Co and Si atoms with the 
tetragonal space group {\sl I4/mcm} isostructural with other $R$Co$_9$Si$_4$. 
The smaller lattice constants of CeCo$_9$Si$_4$ in comparison with the trend established by other $R$Co$_9$Si$_4$ 
is indicative for intermediate valence of cerium.
While $R$Co$_9$Si$_4$ with $R=$ Pr, \dots Tb, and Y show ferromagnetism and LaCo$_9$Si$_4$ is nearly ferromagnetic, 
CeCo$_9$Si$_4$ remains paramagnetic even in external fields as large as 40\,T, though its electronic specific heat 
coefficient ($\gamma\simeq 190$\,mJ/mol\,K$^2$) is of similar magnitude as that of metamagnetic LaCo$_9$Si$_4$ and 
weakly ferromagnetic YCo$_9$Si$_4$.

\end{abstract}

\begin{keyword}

CeCo$_9$Si$_4$ \sep Intermediate Valence \sep High Field Measurement
\end{keyword}

\end{frontmatter}

The ordered ternary rare-earth cobalt silicon phase 
with composition 1:9:4 attracted our attention 
mainly because of the extraordinary magnetic properties of LaCo$_9$Si$_4$ 
which is a strongly exchange enhanced Pauli paramagnet and exhibits an itinerant 
metamagnetic phase transition at about 3.5\,T for 
$H||c$ and 6\,T for $H\bot c$, which is the lowest value ever found for rare earth 
intermetallic compounds~\cite{mi_LCS}. Related isostructural $R$Co$_9$Si$_4$ with 
$R=$ Pr, Nd, Gd, Dy and also Y are ferromagnetic (see Refs.~\cite{huang,mi_YCS}) with 
relatively low $T_C\sim 20-50$\,K. 
This report is on the exceptional behavior of CeCo$_9$Si$_4$ among the ferromagnetic 
(FM) or almost FM $R$Co$_9$Si$_4$ compounds. 

\begin{figure}
\centering
\includegraphics[width=0.96\columnwidth]{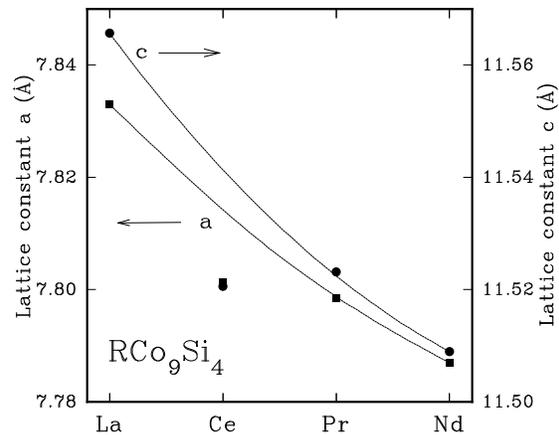}
\caption{Comparison of lattice parameters $a$ (squares) and $c$ (circles) of some 
$R$Co$_9$Si$_4$ compounds; dashed lines guides to the eye.}
\end{figure}
  
Polycrystalline samples of CeCo$_9$Si$_4$ and related $R$Co$_9$Si$_4$ were synthesized by 
induction melting of pure elements ($R$ 3N, Co 4.5N, Si 6N) under protective 
argon atmosphere and subsequent annealing at 1050$^{\circ }$C for one week. 
The crystal structure was determined by means of single crystal X-ray diffraction 
(R$_{F^2}=$\,2.3\%) revealing, analogous to CeNi$_9$Si$_4$~\cite{mi_CNS}, 
a fully ordered distribution of Ce, Co and Si atoms with the LaFe$_9$Si$_4$-type~\cite{tang} 
(space group $I4/mcm$). The lattice parameters are $a=7.801(1)$\,\AA\ and  
$c=11.521(2)$\,\AA\ at room temperature.

\begin{figure}
\centering
\includegraphics[width=0.92\columnwidth]{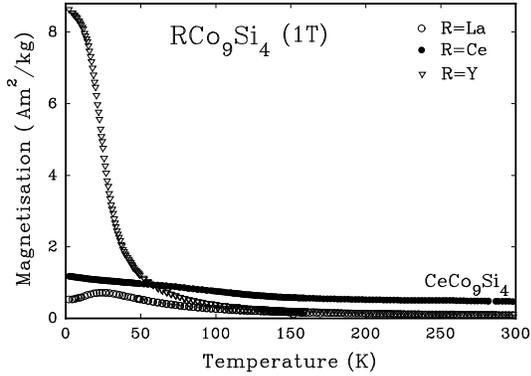}
\caption{Temperature dependent magnetisation $M(T)$ measured at 1T of CeCo$_9$Si$_4$ and 
for comparison LaCo$_9$Si$_4$ and YCo$_9$Si$_4$.}
\end{figure}  
   
The comparison of the lattice parameters of $R$Co$_9$Si$_4$ with $R=$ La, Ce, Pr, Nd 
shown in Fig.~1 reveals a significant negative deviation of the CeCo$_9$Si$_4$ data
from the common trend. 
The smaller lattice constants of CeCo$_9$Si$_4$ indicates an intermediate valence state 
inbetween the magnetic trivalent $4f^1$ and the non-magnetic tetravalent $4f^0$ 
state due to hybridization of $4f$ and conduction band states. 
Intermediate valence close to tetravalence is corroborated
by the magnetisation data shown in Fig.~2 where the Curie-Weiss component 
expected for Ce$^{3+}$ ions is not observed. The dc susceptibility measurement 
in fact reveals an almost temperature independent paramagnetic susceptibility for 
CeCo$_9$Si$_4$, while LaCo$_9$Si$_4$ exhibits a more pronounced temperature 
dependence, typical for a spin fluctuation system, and all other $R$Co$_9$Si$_4$ are 
ferromagnetic below 20--50\,K (see the YCo$_9$Si$_4$ data shown in Fig.~2 as one example). 
The small off-set of the room temperature magnetic moment of CeCo$_9$Si$_4$ as 
compared with LaCo$_9$Si$_4$ and YCo$_9$Si$_4$ in Fig.~2 is due to a contribution from 
traces of unreacted free Co which is also visible in the M(H) data (see Fig.~3). 
The Pauli-susceptibility $\chi_0$ estimated from $M/H$ at 40\,T is about 
$5\times 10^{-7}$\,m$^3$/kg, slightly smaller than $8.2\times 10^{-7}$\,m$^3$/kg 
given in Ref.~\cite{mi_LCS} for LaCo$_9$Si$_4$. 

\begin{figure}
\centering
\includegraphics[width=0.92\columnwidth]{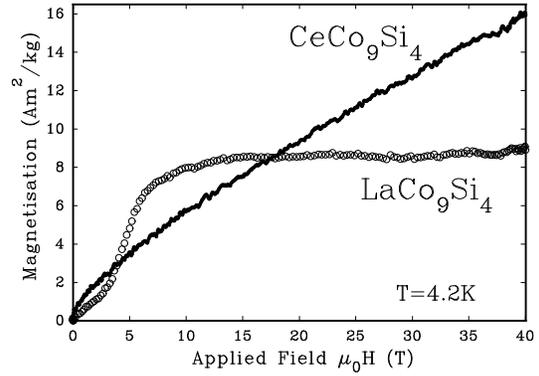}
\caption{High-field magnetisation measurements $M(H)$ on CeCo$_9$Si$_4$ and 
LaCo$_9$Si$_4$ performed at 4.2\,K.}
\end{figure}  
 
Motivated by the metamagnetism observed for LaCo$_9$Si$_4$ high-field magnetisation 
$M(H)$ measurements have been carried out up to 40\,T in pulsed fields (see Ref.\cite{dre} for 
details of the setup) on CeCo$_9$Si$_4$ and LaCo$_9$Si$_4$ for comparison. 
Apart from a small hysteresis at low fields due to the above mentioned Co impurities, 
CeCo$_9$Si$_4$ shows a rather perfectly linear paramagnetic field dependence reaching 
2\,$\mu_B$/f.u.\ at $B_{max}=40$\,T, whereas the magnetic moment of the field induced FM state 
of LaCo$_9$Si$_4$  saturates already above about 8\,T at approximately 1.1\,$\mu_B$/f.u. 

The obviously more stable paramagnetic state of CeCo$_9$Si$_4$ with respect to the appearance of 
metamagnetism is most likely due to $d$-band filling caused by hybridized
$4f$ states of cerium being close to tetravalence. 
Thus, one expects a reduction of spin-fluctuation mass enhancement 
which however may be partly compensated by contributions due to Ce valence fluctuations. 
The latter is supported by the specific heat data (not shown) where the  
linear electronic specific heat coefficient of CeCo$_9$Si$_4$, the Sommerfeld value
$\gamma\simeq 190$\,mJ/molK$^2$, 
is slightly lower than that of nearly magnetic LaCo$_9$Si$_4$, where 
$\gamma\simeq 200$\,mJ/molK$^2$.
The comparison of the temperature dependent resistivities $\rho(T)$, on the other hand
(not shown),   
indicates for both a Fermi liquid behavior with $\rho(T)=\rho_0+AT^2$, however, with a 
significantly lower $A$-coefficient for CeCo$_9$Si$_4$ as compared to 
LaCo$_9$Si$_4$~\cite{CCS}.

We conclude that CeCo$_9$Si$_4$ exhibits intermediate valence with Ce being close to a non-magnetic 
tetravalent state where $4f$ states hybridize with the conduction bands and thereby weaken the magnetic 
correlations among $d$-electrons as compared to other $R$Co$_9$Si$_4$.

\end{document}